\documentclass[11pt,a4paper]{article}

\usepackage{epsfig,amsmath,amssymb,cite}

\def\lp{\left(}
\def\rp{\right)}

\def\rk{\right\}}
\def\lk{\left\{}

\def\be{\begin{equation}}
\def\ee{\end{equation}}
\def\bea{\begin{eqnarray}}
\def\eea{\end{eqnarray}}

\def\p{\partial}

\def\lb{\label}

\def\a{\alpha}
\def\b{\beta}
\def\g{\gamma}
\def\s{\sigma}

\def\om{\omega}

\def\t{\tau}

\def\s{\sigma}
\def\S{\Sigma}
\def\e{\epsilon}
\def\v{\varphi}

\begin{document}

\title{Thin shells joining local cosmic string geometries} 

\author{Ernesto F. Eiroa$^{1,2,}$\thanks{e-mail: eiroa@iafe.uba.ar}, Emilio Rub\'{\i}n de Celis $^{1,3,}$\thanks{e-mail: erdec@df.uba.ar},  
Claudio Simeone$^{1,3,}$\thanks{e-mail: csimeone@df.uba.ar}\\
{\small $^1$ Departamento de F\'{\i}sica, Facultad de Ciencias Exactas y Naturales,} \\ 
{\small Universidad de Buenos Aires, Ciudad Universitaria Pabell\'on I, 1428, Buenos Aires, Argentina}\\
{\small $^2$ Instituto de Astronom\'{\i}a y F\'{\i}sica del Espacio (IAFE, CONICET-UBA),} \\ 
{\small Casilla de Correo 67, Sucursal 28, 1428, Buenos Aires, Argentina}\\
{\small $^3$ IFIBA--CONICET, Ciudad Universitaria Pabell\'on I, 1428, Buenos Aires, Argentina}} 

\maketitle

\begin{abstract}

In this article we present a theoretical construction of spacetimes with a thin shell that joins two different local cosmic string geometries. We study two types of global manifolds, one representing spacetimes with a thin shell surrounding a cosmic string or an empty region with Minkowski metric, and the other corresponding to wormholes which are not symmetric across the throat located at the shell. We analyze the stability of the static configurations under perturbations preserving the cylindrical symmetry. For both types of geometries  we find that the static configurations can be stable for suitable values of the parameters. \\

\noindent 
PACS number(s): 4.20.-q, 04.20.Gz, 04.20.Jb

\end{abstract}

\section{Introduction}
 
Cosmic strings are topological defects which could be the result of symmetry breaking processes in the early Universe \cite{book}.  Their possible important role in the explanation of structure formation at cosmological scale \cite{structure} and, besides, the fact that it would be possible to detect their gravitational lensing effects \cite{lens} led to a considerable amount of work devoted to the study of their associated geometries. From a different point of view, the recent interest in cylindrically symmetric spacetimes was also driven by the present day theoretical framework in which open or closed  strings constitute the most serious basis for a unification of the fundamental interactions (for instance, see Ref. \cite{strings}). Within cosmic string models, particular attention have deserved the so-called gauge or local cosmic strings, which are solutions of the equations of a two-component scalar field coupled to a gauge field. 
The spacetime outside the core of these strings is conical, that is, locally flat but with a deficit angle which is determined by their mass per unit length \cite{cs}. The local flatness of the metric implies zero force on a static particle, but relativistic particles would be deflected, and moving strings would give rise to wakes behind them \cite{structure}. Thus, though cosmic strings are not supposed today to be the main source of the primordial cosmological matter fluctuations, they can still be of relevance as a secondary source of fluctuations \cite{wake}. In the context of axisymmetric spacetimes, the simplicity and the physical interest of the local string metric would make it comparable to the Schwarzschild metric within spherically symmetric geometries.
 
Thin matter layers (``thin shells'') \cite{daris} and their associated geometries appear in both cosmological and astrophysical frameworks. At a cosmological scale, the formalism used to define such layers has been applied in braneworld models, in which a spacetime is defined as the surface  where two higher dimensional manifolds are joined (see for instance \cite{davis} and references therein). At an astrophysical level, such matter layers appear, for example, as models for stellar atmospheres, gravastars, etc. \cite{brady,wilt}. While most thin-shell models studied are associated to spherically symmetric geometries, cylindrical shells have also been considered of interest. In particular, traversable wormhole geometries \cite{motho,whbook,hovi} supported by cylindrical thin shells have been recently studied \cite{cilwh,cms}.  The mechanical stability analysis of thin-shell cylindrical wormholes which are symmetric across the throat was performed in Refs. \cite{mhaeisi}. 
A well known feature of wormholes in general relativity is the presence of exotic matter not satisfying the null energy condition at the throat or in the region close to it \cite{motho,whbook,hovi}. Moreover, it has been shown that the fluid supporting spherically symmetric wormholes is exotic in any scalar-tensor theory of gravity with non-ghost massless scalar field and graviton, and also in $f(R)$ gravity with non-ghost graviton \cite{brsta}. The nature of matter threading cylindrical wormholes was discussed in Refs. \cite{brle,cms}.
Shells related to wormholes, not supporting them but allowing for flat asymptotics, were considered in \cite{bkl}. Static cylindrical shells joining inner and outer regions have been associated to different matter contents and different backgrounds in Refs. \cite{bizoarde}.  Recently, conical spacetimes joined by static shells were the object of field theory calculations in curved space, in the context of a program developed to probe the global aspects of a geometry by the self-force on a charge (see Refs. \cite {eoc}).  
 
Most preceding work deals with different geometries joined by a static shell, or the stability of wormholes which are symmetric across the throat. In the present article we consider shells joining two different conical geometries, that is, the geometries associated to two local cosmic strings with different mass per unit length. We study  the properties of matter on the shells and their mechanical stability under perturbations preserving the symmetry. We analyze both the case of inner solutions joined to outer ones and the case of traversable thin-shell wormholes which are not symmetric across the throat, as those constituting the backgrounds in \cite{eoc}. As usual, we adopt units such that $c=G=1$.

\section{Construction and stability}

We consider static shells and their perturbative stability analysis. Thus we assume a static background geometry, which allows us to define the shells by applying the usual cut and paste procedure starting from two static metrics with cylindrical symmetry representing local cosmic strings. In cylindrical coordinates $x^\alpha_{1,2}=(t_{1,2},r_{1,2},\varphi_{1,2},z_{1,2})$ these metrics can be written in the form
\be \lb{le1}
ds^2_{1,2}=-dt^2_{1,2}+dr^2_{1,2}+r_{1,2}^2 \om_{1,2}^2d\varphi^2_{1,2}+dz^2_{1,2},
\ee
where $0< \om_{1,2} \le 1$ and $0\le \varphi _{1,2} \le 2\pi$. The geometry adopted corresponds to conical spacetimes, which are associated to the so-called gauge or local cosmic strings.  The parameters $\om_{1,2}$ give the deficit angle $2 \pi (1 - \om_{1,2})$ for the corresponding submanifold\footnote{If $\om_{1,2}>1$ the geometry has a surplus angle, case which we do not consider here.}. The deficit angle is proportional to the string mass per unit length, and cannot be removed by a mere redefinition of the angular coordinate; this deficit angle would generate physical effects as double images resulting from the deflection of light, and  matter density fluctuations in the form of wakes in the plane described by the string motion \cite{cs,lens,structure}. 
From these geometries, we take the radii $a_{1,2}>0$ and we construct a geodesically complete manifold $\mathcal{M}=\mathcal{M}_1 \cup \mathcal{M}_2$, which can take any of the following two forms:
\be \lb{nwh}
\mathrm{type\ I:} \quad \mathcal{M}_1=\{x^\alpha_1 / 0\le r_1\le a_1 \}, \qquad \mathcal{M}_2=\{x^\alpha_2 / r_2 \ge a_2 \}; \nonumber
\ee
\be \lb{wh}
\mathrm{type\ II:} \quad \mathcal{M}_1=\{x^\alpha_1 / r_1 \ge a_1 \}, \qquad \mathcal{M}_2=\{x^\alpha_2 / r_2  \ge a_2 \}. \nonumber
\ee
The first case corresponds to joining the interior and the exterior submanifolds while the second case corresponds to joining the two exterior regions\footnote{There is a third case corresponding to joining two interior submanifolds, i.e. $\mathcal{M}_1=\{x^\alpha_1/0\le r_1\le a_1 \} ,\  \mathcal{M}_2=\{x^\alpha_2/0\le r_2\le a_2 \}$; this will not be considered in this work.}.   In both cases, the submanifolds are joined at the hypersurface $\Sigma \equiv \partial \mathcal{M}_1 \equiv \partial \mathcal{M}_2 $ which defines the shell
\be
\S: \left\{\begin{array}{ll}H_1(r_1, \t) = r_1 - a_1(\t) = 0,  \\
H_2(r_2, \t) = r_2 - a_2(\t) = 0, \nonumber
\end{array}\right.
\ee
where for the study of the stability of the configurations, we have let $a_1(\tau)$ and $a_2(\tau)$ to be functions of the proper time $\tau$ on the shell. This proper time can be related to the corresponding coordinate times by 
\be
-d\tau^2=-d{\tau_{1,2}}^2=-{dt_{1,2}}^2+{\dot a_{1,2}}^2d\tau^2,
\ee
where a dot stands for  a derivative with respect to the proper time; thus
\be
d\tau^2=\frac{1}{1+{\dot a_{1,2}}^2}dt^2_{1,2}.
\ee
We adopt the coordinate system $\xi^i=(\tau,\varphi,z)$ on $\S$, with $\varphi=\varphi_1=\varphi_2$ and $z=z_1=z_2$, then the induced metric on the shell is
\be
ds^2 = -d\tau^2 + a_{1,2}^2\om _{1,2}^2 d\varphi ^2 + dz^2.
\ee
The whole spacetime should be continuous across the shell, which is equivalent to require the continuity of the first fundamental form $h_{ij}^{1,2}=\mathrm{diag}(-1,a_{1,2}^2\om _{1,2}^2,1)$ on $\S$, so we find the following equation
\be \lb{fff} 
a_1 \omega_1 = a_2 \omega_2,
\ee
that relates the radial coordinate of the shell as seen from both sides of it. 

While the geometry must be continuous across the joining surface, the derivatives of the metric are not forced to such restriction. In general, there can be a jump in these derivatives which is associated with the presence of a thin layer of matter. The covariant form of this relation between derivatives of the metric and the matter on the shell are the Lanczos equations \cite{daris}
\be  \lb{le2}
8\pi S_{ij}  = [K] h_{ij} - [K_{ij} ],
\ee
where $[K_{ij}] = K_{ij}^{2} - K_{ij}^{1}$ is the discontinuity of the extrinsic curvature tensor across the shell, $[K] = h^{ij} [K_{ij}]$ is the corresponding trace, and $S_{ij}$ is the surface stress-energy tensor. The extrinsic curvature tensor is given by
\be
K^{1,2}_{ij} = -n^{1,2}_{\g} \left( \frac{\p^2 x^{\g}_{1,2}}{\p \xi^i \p \xi^j} + (\Gamma_{1,2})^{\g}_{\a \b} \frac{\p x^{\a}_{1,2}}{\p \xi^i} \frac{\p x^{\b}_{1,2}}{\p \xi^j} \right) \Bigg|_{\S}
\ee
where the unit normals to the surface $\S$ pointing from $\mathcal{M}_1$ to $\mathcal{M}_2$ are
\be
n^{1,2}_{\g} = \delta  \, \Bigg| g^{\a\b}_{1,2} \frac{\p H_{1,2}}{\p x^{\a}_{1,2}} \frac{\p H_{1,2}}{\p x^{\b}_{1,2}} \Bigg|^{-1/2} \frac{\p H_{1,2}}{\p x^{\g}_{1,2}} ,
\ee
with $\delta$ given by
\be
\delta  = \lk \begin{array}{ll}
\;\;+1 , \quad & \mbox{type I geometry},\\
\left.\begin{array}{ll}
-1 ,\; \mathrm{for}\, \mathcal{M}_1 \\
+1 ,\; \mathrm{for}\, \mathcal{M}_2
\end{array} \rk \, & \mbox{type II geometry}. \nonumber
\end{array}\right.
\ee
The explicit calculation in our case yields the normal 4-vector
\be
n^{1,2}_\g= \delta  \left( -{\dot a_{1,2}}, \sqrt{1+{\dot a_{1,2}}^2}, 0, 0\right).
\ee
In the orthonormal basis on the shell we have that $h_{\hat i \hat j}=\mathrm{diag}(-1,1,1)$ and the non-vanishing components of the extrinsic curvature tensor are given by
\be
K^{1,2}_{\hat \tau \hat \tau} = - \delta  \frac{\ddot a_{1,2} }{\sqrt{1+{\dot a_{1,2}}^2}},
\qquad
K^{1,2}_{\hat \varphi\hat \varphi}= \delta  \frac{\sqrt{1+{\dot a_{1,2}}^2}}{a_{1,2}},
\qquad
\mathrm{and}
\qquad
K^{1,2}_{\hat z\hat z} = 0.
\ee
Finally, for the jump $[ K_{\hat{i} \hat{j}} ]$ across the shell we obtain
\be
[K_{\hat \t \hat \t}] = -\frac{\ddot a_2}{\sqrt{1+{\dot a_2}^2}}  -\e  \frac{\ddot a_1}{\sqrt{1+{\dot a_1}^2}},
\qquad
[K_{\hat \varphi\hat \varphi}] = \frac{\sqrt{1+{\dot a_2}^2}}{a_2} +\e  \frac{\sqrt{1+{\dot a_1}^2}}{a_1},
\qquad
\mathrm{and}
\qquad
[K_{\hat z\hat z}] =0,
\ee
where $\e  =-1$ corresponds to a type I geometry and $\e  =1$  to a type II spacetime. In the orthonormal frame, the surface stress-energy tensor has the form $S_{\hat i \hat j} = \mathrm{diag}(\s, p_\varphi,p_z) $, with $\s$ the surface energy density, and $p_\v$ and $p_z$ the surface pressures. From Eq. (\ref{le2}), we find that the energy density can be written as
\be  \lb{sig}
\s  =  -\frac{\sqrt{1+{\dot a_2}^2}}{8\pi a_2} - \e \frac{\sqrt{1+{\dot a_1}^2}}{8\pi a_1};
\ee
analogously, the pressures take the form
\be \lb{presphi}
p_\varphi=\frac{\ddot a_2} {8 \pi \sqrt{1+{\dot a_2}^2}}  + \e   \frac{\ddot a_1} {8 \pi \sqrt{1+{\dot a_1}^2}} 
\ee
and
\be \lb{presz}
p_z=\frac{ 1+{\dot a_2}^2 + a_2\ddot a_2}{8\pi a_2\sqrt{1+{\dot a_2}^2}}  + \e \frac{ 1+{\dot a_1}^2 + a_1\ddot a_1}{8\pi a_1\sqrt{1+{\dot a_1}^2}}.
\ee
It is important to note that the energy density and the pressures are the result of the sum of two contributions coming from each side of the shell, with the corresponding radii of the shell and their time derivatives related by Eq. (\ref{fff}). From these equations it is useful to compute the static values of the energy density and pressures which will be used in subsequent calculations. For the static energy density we have
\be \lb{s0}
\s_0 =   -\frac{1}{8 \pi a_{20}} - \e  \frac{1}{8 \pi a_{10}} ,
\ee
while for the pressures we obtain
\be \lb{p0f}
p_{0 \v} = 0
\ee
and
\be \lb{p0z}
p_{0z} = \frac{1}{8 \pi a_{20}} + \e  \frac{1}{8 \pi a_{10}} ,
\ee
where $a_{10}$ and $a_{20}$ are the static values of the radial coordinate of the shell corresponding to each side, linked by Eq. (\ref{fff}), i.e. $a_{10} \om_1 = a_{20}\om_2$.

In principle, if the equations of state relating the pressures with the energy density are provided, the equations  (\ref{sig}),  (\ref{presphi}), and  (\ref{presz}) above can be integrated to obtain the time evolution of the position of the shell. We can write the energy density and the pressures (given by Eqs. (\ref{sig}), (\ref{presphi}), and (\ref{presz})) as functions only of the radial coordinate $a_2\equiv a$ and its time derivatives, by using Eq. (\ref{fff}), so that $a_1=a \om_2 / \om_1$. Every expression that follows is understood to depend on $a$. By taking the proper time derivative in Eq. (\ref{sig}) and using Eqs. (\ref{presphi}) and (\ref{presz}), we obtain the conservation equation in the form
\be \lb{cons}
\dot \s + \s \frac{\dot a}{a}  + p_{\v} \frac{\dot a}{a} = 0.
\ee
This conservation equation can be rewritten to give the differential equation
\be \lb{difsigma2}
a \s' + \s  + p_{\v} = 0,
\ee
where the prime represents the derivative with respect to $a$. By introducing the equation of state relating the pressure $p_\v$ and the energy density, the first order differential equation (\ref{difsigma2}) can be integrated to  obtain $\s (a)$ from
\be
\int_{\s_0}^\s\frac{d\s}{\s+p_\v(\s)}=-\int_{a_0}^a\frac{da}{a}.
\ee
Here we are only interested in the stability of the static configurations and not in the dynamical evolution of the shell. In what follows, we assume linearized equations of state around the static configurations for both pressures:
\be
p_{\v} = \eta_{\v} \lp \s-\s_0 \rp + p_{0\v}
\ee
and
\be
p_{z } = \eta_z  \lp \s-\s_0 \rp + p_{0z},
\ee
where $\eta_{\v}$ and $\eta_z$ are constants. Then, from Eq. (\ref{sig}) we can write the equation of motion for the shell in the form
\be
\dot{a}^2 + V(a) = 0 ,
\ee
with the potential
\be \lb{pot}
V(a)=\frac{\om_1^2+\om_2^2}{2\om_2^2} - \lp \frac{\om_1^2-\om_2^2}{16\pi a \om_2 \s} \rp ^2  - (4\pi a\s)^2,
\ee
written in terms of the energy density $\s (a)$. A second order Taylor expansion of this potential around the static solution found at $a_0$ gives
\be
V(a) = V (a_0) + V'(a_0) (a-a_0) + V''(a_0) (a-a_0)^2 + O(a-a_0)^3.
\ee
From (\ref{pot}) it is easy to check that $V(a_0)$=0. The first derivative of the potential, using the conservation equation written in the form  (\ref{difsigma2}), gives $V'(a_0)=0$. Finally, introducing the equations of state and using (\ref{difsigma2}) again, the second derivative of the potential evaluated at the equilibrium position reads
\be \lb{potd2}
V''(a_0) = -\frac{2 \e \eta_{\v } \om_1 }{a_0^2 \om_2 } .
\ee
The static solution is stable under radial perturbations if $V''(a_0)>0$. This condition, solves the general problem of the mechanical linearized stability for both type I and type II topologies.

\section{Examples}

In this section, we apply the formalism introduced above to examples: we construct a thin shell surrounding a local cosmic string and a cylindrical wormhole not symmetric across the throat corresponding to the shell.

\subsection{Thin shells}

Firstly, we present an application of the formalism in the case of a type I geometry, i.e. we construct a spacetime in which an interior region is joined by means of a thin shell to an exterior one which extends to infinity in the radial coordinate. For our construction we adopt two geometries of local cosmic string form with different angle deficits. When these two submanifolds are glued, the resulting geometry is that corresponding to a shell surrounding a local cosmic string.  Using Eqs. (\ref{s0}), (\ref{p0f}), and (\ref{p0z}), the static values of the energy density and the pressures are, respectively, 
\be \lb{s0LC1}
\s_0 = \frac{1}{8 \pi a_0} \frac{\om_1-\om_2}{ \om_2} , 
\ee
\be
p_{0 \v} = 0,
\ee
and
\be
p_{0 z} = \frac{1}{8 \pi a_0} \frac{\om_2 -\om_1}{\om_2}.
\ee
We see that these magnitudes can have negative or positive values. In particular, Eq. (\ref{s0LC1}) shows that exotic matter (i.e. not satisfying the weak and null energy conditions) or ordinary matter is needed to construct the shell depending on the value of the factor $\om_1 - \om_2$: if the deficit angle of the outer region is larger than that of the inner region, the energy density is positive, while in the reverse situation the energy density can only be negative, then corresponding to a shell of exotic matter. We also find that for conical spacetimes the  pressure $p_\v$ of a static shell vanishes, and the tension  along  the axis of symmetry equals the energy density; this could be expected  from the nature of the stress-energy tensor associated to the cosmic string static geometries from which the mathematical construction starts (see \cite{cs}). 
In the particular case that $\om_1=1$, the inner region has a Minkowski metric, and the conical singularity at the axis of symmetry does not exist. Instead, if $\om_2=1$, we have a Minkowski geometry in the outer region, which then presents no angle deficit. The shell in this case does a sort of screening of the cosmic string inside. 
The potential for the equation of motion of the thin shell is given by Eq. (\ref{pot}), and its second derivative evaluated at the static position $a_0$ is obtained from Eq. (\ref{potd2}), with $\e =-1$. The sign of $V''(a_0)$ is independent of the parameters $\om_1$ and $\om_2$, because these quantities are always positive, and also from $a_0$. 
This means that the stability only depends on the parameters of the linearized equations of state. That is, in the case of local cosmic string submanifolds, the stability result can be straightforwardly read from the expression of the second derivative of the potential: it reduces to the condition of a positive coefficient $\eta_{\v}$ in the associated linearized equation of state.

\subsection{Wormholes}

As a second application of the formalism, we construct cylindrical thin--shell wormholes which are not symmetric across the throat. We then take  a type II geometry, with the shell that corresponds to the throat if the flare--out condition is satisfied. In the literature there are two definitions of this condition for cylindrical geometries. In the first one, called the {\it areal} condition, the area functions ${\cal A}_{1,2}(r_{1,2})=2\pi \sqrt{g_{\varphi\varphi}^{1,2}(r_{1,2})g_{zz}^{1,2}(r_{1,2})}= 2\pi \om_{1,2}r_{1,2}$ should increase at both sides of the throat. The second one \cite{brle}, known as the  {\it radial} condition, only requires that the circular radius functions ${\cal R}_{1,2}(r_{1,2}) = \sqrt{g_{\varphi\varphi}^{1,2}(r_{1,2})} = \om_{1,2}r_{1,2}$ have a minimum at the throat.  It is clear that the two possible flare-out conditions, which in our construction only differ by a constant factor, are satisfied because $r_{1,2} \ge a_{1,2}$. 
This mathematical construction creates a complete manifold with two different regions at the sides of the throat corresponding to the conic submanifolds $\mathcal{M}_1$ and $\mathcal{M}_2$, determined by Eq. (\ref{wh}), which are joined to form a thin-shell wormhole. From Eqs. (\ref{s0}), (\ref{p0f}), and (\ref{p0z}), the static values of the energy density and pressures at the throat are
\be \lb{s0wh}
\s_0 = -\frac{1}{8 \pi a_0} \frac{\om_1+\om_2}{ \om_2} ,
\ee
\be
p_{0 \v} = 0,
\ee
and
\be
p_{0 z} = \frac{1}{8 \pi a_0} \frac{\om_2 + \om_1}{\om_2},
\ee
which corresponds to negative energy density and positive pressure along the symmetry axis for the local cosmic string wormholes. The matter at the shell is always exotic. As in the preceding subsection,  we obtain that for the joining of two conical submanifolds the pressure $p_\v$ of the static configuration vanishes and the negative tension along the axis equals the energy density, as we could expect for the same reason pointed out in the preceding section. In the case that $\om_1=1$ or $\om_2=1$ the corresponding side of the wormhole has a Minkowski geometry. 
The potential for the wormholes has the same form of the previous subsection, i.e. Eq. (\ref{pot}) with the corresponding energy density $\s$. The second derivative of the potential evaluated at the static position is given by Eq. (\ref{potd2}), now with $\e =1$. In the case of wormholes which connect two local cosmic string spacetimes, from the sign of this second derivative the stability condition $\eta_{\v}<0$ immediately becomes apparent.

\section{Summary}

We have presented the matter characterization and the stability analysis of cylindrical shells joining two local cosmic string metrics. We have studied two types of global geometries, the first one associated to spacetimes with a thin shell separating an inner from an outer region, and the second one describing wormholes with a shell at the throat connecting two different geometries.  In the case of wormholes, the fluid at the throat is always exotic, i.e. it does not satisfy the null and weak energy conditions. In the case of shells where an inner region joins an outer one, the matter can be ordinary or exotic, depending on the relation between the corresponding angle deficits. 
For the stability analysis of static configurations, we have adopted linearized equations of state that relate the pressures $p_\varphi$ and $p_z$  with the surface energy density $\sigma$ at the shell, with  coefficients $\eta_\varphi$ and $\eta_z$, respectively. 
We have found that the static configurations are stable if suitable values of the parameters are adopted, for both shells and wormholes.  The pressure along the axis turns out to be not relevant for stability, and only the coefficient associated to the angular pressure enters the stability condition; this is reasonable for  radial perturbations, i.e. perturbations which involve a change of the circumference perimeter are expected to be affected by forces in the angular direction. For wormholes to be stable the coefficient $\eta _\varphi $ must always be negative; instead, in the case of shells separating an interior region from an exterior one, we find the desirable feature that they can be stable when $\eta _\varphi >0$.

\section*{Acknowledgments}

This work has been supported by CONICET and Universidad de Buenos Aires.


\begin{thebibliography}{99}
 
\bibitem{book} T. W. B. Kibble, Phys. Rept. \textbf{67}, 183 (1980); A. Vilenkin and E. P. S. Shellard, \textit{Cosmic Strings and Other Topological Defects} (Cambridge University Press, Cambridge, 1994).
 
\bibitem{structure} A. Vilenkin, Phys. Rev. Lett. \textbf{46}, 1169 (1981); \textbf{46}, 1496(E) (1981); N. Turok and R. H. Brandenberger, Phys. Rev. D \textbf{33}, 2175 (1986); H. Sato,  Prog. Theor. Phys. \textbf{75}, 1342 (1986); T. Vachaspati, Phys. Rev. Lett. \textbf{57}, 1655 (1986); T. Vachaspati and A. Vilenkin, Phys. Rev. Lett. \textbf{67},  1057 (1991).
 
\bibitem{lens} A. Vilenkin, Astrophys. J. 282, L51 (1984); J. R. Gott III, Astrophys. J. \textbf{288},  422 (1985).
 
\bibitem{strings} A. Vilenkin, in \textit{Inflating Horizons of Particle Astrophysics and Cosmology}, edited by H. Suzuki, J. Yokoyama, Y. Suto, and K. Sato (Universal Academy Press, Tokyo, 2006) (arXiv:hep-th/0508135).

\bibitem{cs} A. Vilenkin,  Phys. Rev. D \textbf{23},  852 (1981); W. A. Hiscock, Phys. Rev. D \textbf{31}, 3288 (1985).

\bibitem{wake}J . Magueijo, A. Albrecht, D. Coulson and P. Ferreira, Phys. Rev. Lett. \textbf{76}, 2617 (1996); U. L. Pen, U. Seljak and N. Turok,  Phys. Rev. Lett. \textbf{79}, 1611 (1997); A. Nayeri, R. H. Brandenberger and C. Vafa, Phys. Rev. Lett. \textbf{97}, 021302 (2006); R. J. Danos, R. H. Brandenberger, and G. Holder, Phys. Rev. D \textbf{82}, 023513 (2010).

\bibitem{daris} N. Sen, Ann. Phys. (Leipzig) \textbf{378}, 365 (1924); K. Lanczos, Ann. Phys. (Leipzig)  \textbf{379}, 518 (1924); G. Darmois, M\'{e}morial des Sciences Math\'{e}matiques, Fascicule XXV  (Gauthier-Villars, Paris, 1927), Chapter 5; W. Israel, Nuovo Cimento B \textbf{44}, 1 (1966); \textbf{48}, 463(E) (1967).
 
\bibitem{davis} S. C. Davis, Phys. Rev. D {\bf 67}, 024030 (2003).
 
\bibitem{brady} P. R. Brady, J. Louko, and E. Poisson, Phys. Rev. D {\bf 44}, 1891 (1991).  
 
\bibitem{wilt} M. Visser and D. L. Wiltshire, Class. Quantum Grav. {\bf 21}, 1135 (2004).
 
\bibitem{motho} M. S. Morris and K. S. Thorne, Am. J. Phys. \textbf{56}, 395 (1988); M. S. Morris, K. S. Thorne and U. Yurtsever, Phys. Rev. Lett {\bf 61}, 1446 (1988); V. P. Frolov and I. D. Novikov, Phys. Rev. D {\bf 42}, 1057 (1990).
 
\bibitem{whbook} M. Visser, \textit{Lorentzian Wormholes} (AIP Press, New York, 1996).

\bibitem{hovi} D. Hochberg and M. Visser, Phys. Rev. D \textbf{56}, 4745 (1997); M. Visser, S. Kar and N. Dadhich, Phys. Rev. Lett. \textbf{90}, 201102 (2003).
 
\bibitem{cilwh}  E. F. Eiroa and C. Simeone, Phys. Rev. D \textbf{70}, 044008 (2004); C. Bejarano,  E. F. Eiroa, and C. Simeone, Phys. Rev. D \textbf{75}, 027501 (2007); M. G. Richarte and C. Simeone, Phys. Rev. D \textbf{79}, 127502 (2009); E. F. Eiroa and C. Simeone, Phys. Rev. D \textbf{81}, 084022 (2010); \textbf{90}, 089906(E) (2014); M. Sharif and M. Azam,  JCAP \textbf{1304}, 023 (2013); M. G. Richarte, Phys. Rev. D \textbf{88}, 027507 (2013); S. Habib Mazharimousavi, M. Halilsoy, and Z. Amirabi, Eur. Phys. J. C \textbf{74}, 2889 (2014).  

\bibitem{cms} C. Simeone, Int. J. Mod. Phys. D \textbf{21}, 1250015 (2012).

\bibitem{mhaeisi} S. Habib Mazharimousavi, M. Halilsoy, and Z. Amirabi,  Phys. Rev. D \textbf{89}, 084003 (2014); E. F. Eiroa and C. Simeone, Phys. Rev. D  \textbf{91}, 064005 (2015).
 
\bibitem{brsta} K. A. Bronnikov and A. A. Starobinsky, Mod. Phys. Lett. A \textbf{24}, 1559 (2009); K. A. Bronnikov, M. V. Skvortsova and A. A. Starobinsky, Grav. Cosmol. \textbf{16}, 216 (2010). 
 
\bibitem{brle} K. A. Bronnikov and J. P. S. Lemos, Phys. Rev. D \textbf{79}, 104019 (2009).
   
\bibitem{bkl} K. A. Bronnikov, V. G. Krechet and J. P. S. Lemos, Phys. Rev. D \textbf{87}, 084060 (2013).

\bibitem{bizoarde} J. Bicak and M. Zofka, Class. Quantum Grav. {\bf 19}, 3653 (2002); M. Arik and O. Delice, Gen. Relativ. Gravit. {\bf 37}, 1395 (2005); M. Zofka and J. Bicak, Class. Quantum Grav. {\bf 25}, 015011 (2008).

\bibitem{eoc} E. Rub\'{\i}n de Celis, O. Santill\'an and C. Simeone, Phys. Rev. D \textbf{86}, 124009 (2012); E. Rub\'{\i}n de Celis, Eur. Phys. J. C  \textbf{76}, 92 (2016).

\end{thebibliography}
\end{document}